\documentclass[useAMS,usenatbib]{mn2e}
\usepackage{amsmath}
\usepackage{graphicx, subfigure}
\usepackage{color}
\usepackage{ulem}
\usepackage{caption}

\def\be{\begin{equation}} 
\def\ee{\end{equation}}

\def\mhh{{$M_{\rm H_2}$}}

\def\HH{${\rm {H_2}}$}

\def\HI{\hbox{H~$\scriptstyle\rm I\ $}}

\def\gsim{\lower.5ex\hbox{\gtsima}} 
\def\lsim{\lower.5ex\hbox{\ltsima}} 
\def\gtsima{$\; \buildrel > \over \sim \;$} 
\def\ltsima{$\; \buildrel < \over \sim \;$} 
\def\prosima{$\; \buildrel \propto \over \sim \;$} \def\gsim{\lower.5ex\hbox{\gtsima}} 
\def\lsim{\lower.5ex\hbox{\ltsima}} 
\def\simgt{\lower.5ex\hbox{\gtsima}} 
\def\simlt{\lower.5ex\hbox{\ltsima}} 
\def\simpr{\lower.5ex\hbox{\prosima}} \def\la{\lsim} \def\ga{\gsim} 
  
\def\gtsima{$\; \buildrel > \over \sim \;$} 
\def\ltsima{$\; \buildrel < \over \sim \;$} 
\def\gsim{\lower.5ex\hbox{\gtsima}} 
\def\lsim{\lower.5ex\hbox{\ltsima}} 
\def\simgt{\lower.5ex\hbox{\gtsima}} 
\def\simlt{\lower.5ex\hbox{\ltsima}} 
\def\simpr{\lower.5ex\hbox{\prosima}} 
\def\la{\lsim} 
\def\ga{\gsim}

\def\E3{{\cal E}_{\rm g}^{III}}

\title[Molecular hydrogen in LAEs]{Molecular hydrogen in Lyman Alpha Emitters}
\author[Vallini, Dayal \& Ferrara]{Livia Vallini$^{1}$\thanks{E-mail: 
livia.vallini@sns.it (LV)}, Pratika Dayal$^{2,3}$,  \& Andrea Ferrara$^{1}$\\ 
$^{1}$ Scuola Normale Superiore, Piazza dei Cavalieri 7, 56126 Pisa, Italy \\
$^{2}$ Institute for Astronomy, The University of Edinburgh, Royal Observatory, Blackford Hill, Edinburgh, EH9 3HJ, UK \\
$^{3}$ Leibniz-Institute for Astrophysics, Potsdam, An der Sternwarte 16, Potsdam, Germany, 14482 }
\begin{document}


\pagerange{\pageref{firstpage}--\pageref{lastpage}} \pubyear{2011}

\maketitle

\label{firstpage}

\begin{abstract}
We present a physically motivated model to estimate the molecular hydrogen (\HH) content of high-redshift ($z \approx 5.7,6.6$) Lyman Alpha Emitters (LAEs) extracted from a suite of cosmological simulations. We find that the $\rm{H_2}$ mass fraction, $f_{\rm{H_2}}$, depends on three main LAE physical properties: (a) star formation rate, (b) dust mass, and (c) cold neutral gas mass. At $z \approx 5.7$, the value of $f_{\rm H_2}$ peaks and ranges between $0.5-0.9$ for intermediate mass LAEs with stellar mass $M_* \approx 10^{9-10} M_\odot$, decreasing for both smaller and larger galaxies. However, the largest value 
of the $\rm{H_2}$ mass is found in the most luminous LAEs. These trends also hold at $z \approx 6.6$, although, due to a lower dust content, $f_{\rm{H_2}}(z=6.6)  
 \approx 0.5 f_{\rm{H_2}}(z=5.7)$  when averaged over all LAEs; they arise due to the interplay between the $\rm{H_2}$ formation/shielding controlled by dust and the intensity of the ultraviolet (UV) Lyman-Werner photo-dissociating radiation produced by stars. We then predict the carbon monoxide (CO) luminosities for such LAEs and check that they are consistent with the upper limits found by Wagg et al. (2009) for two $z >6$ LAEs. At $z \approx 5.7, 6.6$, the lowest CO rotational transition observable for both samples with the actual capabilities of Atacama Large Millimeter Array (ALMA) is the CO(6-5). We find that at $z \approx 5.7$, about 1-2\% of LAEs, i.e., those with an observed Lyman Alpha luminosity larger than $10^{43.2} {\rm erg \, s^{-1}}$ would be detectable with an integration time of 5-10 hours ($S/N=5$); at $z \approx 6.6$ none of the LAEs would be detectable in CO, even with an ALMA integration time of 10 hours. We also build the CO `flux function', i.e., the number density of LAEs as a function of the line-integrated CO flux, $S_{\rm{CO}}$, and show that it peaks at $S_{\rm{CO}} = 0.1$ mJy at $z=5.7$, progressively shifting to lower values at higher redshifts. We end by discussing the model uncertainties.
\end{abstract}


\section{Introduction}
\label{intro}
The past few years have seen a rapid increase in the amount of data available on high-redshift galaxies. This has been made possible by a combination of state of the art instruments such as the Hubble Space Telescope, the Subaru and Keck telescopes, and sophisticated selection techniques. Of the latter, one of the most successful approaches has been the use of the narrow-band technique \citep[e.g.][]{malhotra2005, shimasaku2006, kashikawa2006, hu2010} that is based on looking for the Lyman alpha (Ly$\alpha$) emission at 1216 \AA\, in the galaxy rest frame. Hundreds of such galaxies, called LAEs have now been confirmed at $z\approx2. 25$ \citep{nilsson2009}, $z\approx3$ \citep{cowie1998, steidel2000, matsuda2005, venemans2007, ouchi2008}, $z\approx4.5$ \citep{finkelstein2007}, $z\approx5.7$ \citep{malhotra2005,shimasaku2006} $z\approx6.6$ \citep{taniguchi2005, kashikawa2011} and $z\approx7$ \citep{iye2006}. Due to their large number statistics and unambiguous spectral signatures, LAEs are arguably the best probes of reionization and high-redshift galaxy evolution \citep{santos2004, dijkstra2007, kobayashi2007, kobayashi2010, dayal2008, dayal2009, dayal2010a, dayal2011}. 

However, using LAEs as probes of high-redshift galaxy populations, understanding them to study reionization, and calculating their contribution to reionization requires an understanding of their star formation rates (SFR). Translating the observed UV luminosity (1375 \AA\, in the galaxy rest frame) into an intrinsic SFR is rendered hard by the fact that the continuum photons produced in a galaxy are attenuated by the dust in the galactic interstellar medium (ISM) before they reach the observer. Inferring the intrinsic SFR using the observed Ly$\alpha$ luminosity is even more complicated since Ly$\alpha$ photons are absorbed both by the ISM dust, as well as the neutral hydrogen (\HI) in the intergalactic medium (IGM) along the line of sight between the emitter and the observer. Indeed, a number of studies point to LAEs being dust enriched, even at redshifts $z \approx 6$: using theoretical models, \citet{dayal2010a} have shown that at $z \approx 5.7$, the color excess of LAEs, $E(B-V) \sim 0.15$ while observationally, the color excess values range between $E(B-V) \approx 0.025-0.32$ at $z \approx 4-5.7$ \citep{lai2007, pirzkal2007, finkelstein2009}. Further, the observed Ly$\alpha$ luminosity depends both on the reionization state of the IGM, as well as on the IGM peculiar velocities along the line of sight; inflows/outflows into/from a galaxy can blueshift/redshift the Ly$\alpha$ line, thereby decreasing/increasing the IGM Ly$\alpha$ transmission, $T_\alpha$ \citep{verhamme2006, dayal2011b, dijkstra2011}. However, the extent to which peculiar velocities influence $T_\alpha$ is debatable since these calculations have mostly been performed under idealized situations. For example, \citet{verhamme2006} have used spherically symmetric outflows of \HI to show an enhancement in $T_\alpha$; however, many studies, e.g. \citet{fangano2007} and references therein, have shown that Kelvin-Helmholtz instabilities would result in breaking-up such symmetric outflows. 

In this sense, $\rm{H_2}$ is a far better indicator of the SFR since stars form in dense, cold molecular clouds (MCs); theoretical and observational constraints on the latter are then of utmost importance to shed light on the molecular content, and therefore the intrinsic SFR of these high-redshift galaxies. The $\rm{H_2}$ content of galaxies is generally studied through observations of CO rotational emission lines \citep{solomon2005,omont2007} that have been detected in more than a hundred high-redshift sources, even though searches for molecular gas at redshifts $z>4$ have so far focused mainly on quasars and on the most massive, far-infrared-luminous and submillimeter galaxies \citep[see][]{riechers2011}. As of now, only scant effort has been devoted to observing the molecular content of high-redshift ($z \geq 6$) LAEs. In one of the only observational works available, \citet{wagg2009} have searched for low-J rotational CO emission lines in two LAEs at $z>6.5$; the non-detection of any CO emission from these galaxies can then be used to put constraints on the amount of molecular gas in these sources. 

In this work, our aim is to present a self-consistent and physically motivated model to calculate the ${\rm H_2}$ fraction and mass, and relate it to the physical properties of the galaxies identified as LAEs at $z \approx 5.7, 6.6$. To do so, we start by using a previously developed LAE model \citep[see][]{dayal2008, dayal2009, dayal2010a,dayal2011}, where the authors combined state of the art cosmological SPH simulations with a Ly$\alpha$ production/transmission model to successfully reproduce a large number of observational data sets. We couple this with a semi-analytic model that describes the structure of the MCs, to calculate the molecular hydrogen content taking into account its formation on dust grains, its destruction by UV photons, and the shielding by \HI in the ISM, as proposed by \citet{krumholz2008, krumholz2009} and \citet{mckee2010}. Once the molecular fraction is calculated for all of the LAEs at $z \approx 5.7, 6.6$, we examine its correlations with the physical properties of the emitters, including the SFR, the total ISM dust mass and the amount of cold \HI gas in the ISM. Translating the total ${\rm H_2}$ mass into a CO luminosity, we compare the predictions of our model to the observations of \citet{wagg2009}. Finally, we estimate the time required to observe such CO emission with ALMA, one of whose main goals is to observe molecular gas in high-$z$ sources. 

\section{The cosmological simulations}
\label{simulations}

In this Section, we briefly describe the simulation used in this work and interested readers are referred to \citet{tornatore2010} for a complete description. The simulation has been carried out using the TreePM-SPH code GADGET-2 \citep{springel2005} with the implementation of chemodynamics as described in \citet{tornatore2007}. The adopted cosmological model corresponds to a $\rm{\Lambda CDM}$ Universe with $\Omega_m=0.26$, $\Omega_{\Lambda} = 0.74$, $\Omega_ b = 0.0413$, $n_s = 0.95$, $\rm{H_0} = 73\,\rm{km\,s^{-1}\,Mpc^{-1}}$ and $\sigma_{8}=0.8$, thus consistent with the 5-year analysis of the WMAP data \citep{komatsu2009}. The simulation has a periodic box size of $75h^{-1}\,$ comoving Mpc (cMpc) and contains $512^3$ dark matter (DM) particles, and initially the same number of baryonic particles. The run assumes a uniform redshift-dependent UV background produced by quasars and galaxies, as given by \citet{haardt1996} and includes metallicity-dependent radiative cooling \citep{sutherland1993}. The code has an effective model to describe star formation from a multi-phase ISM and a prescription for galactic winds triggered by supernova (SN) explosions \citep{springel2003}; the initial mass function (IMF) is taken to be Salpeter between $1-100\,\rm{M_{\odot}}$. Metals and energy are released by stars of different masses by properly accounting for mass-dependent lifetimes as proposed by \citet{padovani1993}. The code uses the metallicity-dependent yields from \citet{woosley1995}; the yields for SNIa and asymptotic giant branch (AGB) stars have been taken from \citet{vandenhoek1997}.

Galaxies are identified as gravitationally bound groups of star particles by running a friends-of-friends (FOF) algorithm. Each FOF group is then decomposed into a set of disjoint substructures, which are identified as locally overdense regions in the density field of the background main halo by the SUBFIND algorithm \citep{springel2001}.  After performing a gravitational unbinding procedure, only sub-halos with at least 20 bound particles are considered to be genuine structures \citep{saro2006}. For each galaxy in each of the snapshots of interest ($z \approx 5.7,6.6$), we obtain the total halo/gas/stellar mass ($M_h/M_g/M_*$), the SFR ($\dot M_*$), the mass weighted gas/stellar metallicity ($Z_g/Z_*$), the mass weighted age ($t_*$), the mass weighted gas temperature and the half mass radius of the dark matter halo. 

\subsection{Identifying LAEs}
\label{identify_laes}

The simulated properties of each galaxy at $z \approx 5.7,6.6$ are used to calculate the total intrinsic Ly$\alpha$ ($L_\alpha^{int}$) and continuum luminosity ($L_c^{int}$) which include both the contribution from stellar sources and from the cooling of collisionally excited \HI in the ISM as shown in \citet{dayal2010a}. To calculate the stellar contribution, we obtain the spectrum of each LAE using the population synthesis code {\tt STARBURST99} \citep{leitherer1999}, using the simulated values of $t_*, M_*, \dot M_*$; the cooling radiation depends on the temperature of the ISM gas. The intrinsic Ly$\alpha$ luminosity can be translated into the observed luminosity such that $L_\alpha = L_\alpha^{int} f_\alpha T_\alpha$, while the observed continuum luminosity, $L_c$ is expressed as $L_c = L_c^{int} f_c$. Here, $f_\alpha$ ($f_c$) are the fractions of Ly$\alpha$ (continuum) photons escaping the galaxy, undamped by the ISM dust and $T_\alpha$ is the fraction of the Ly$\alpha$ luminosity that is transmitted through the IGM, undamped by neutral hydrogen.

The main features of the model used to calculate $f_c, f_\alpha$ and $T_\alpha$ are: (a) for each galaxy the dust enrichment is derived by using its 
intrinsic properties ($\dot M_*$, $t_*$, $M_g$) and assuming Type II supernovae (SNII) to be the primary dust factories. The dust mass, $M_d$, is calculated 
including dust production due to SNII, assuming an average dust mass produced per SNII of $0.5\,\rm{M_\odot}$
\citep[][]{todini2001, nozawa2003, nozawa2007, bianchi2007}, dust destruction with an efficiency of about 40\% in the region shocked to 
speeds $\geq 100~{\rm km \, s^{-1}}$ by SNII shocks, assimilation of a homogeneous mixture of dust and gas into subsequent SF (astration), and 
ejection of a homogeneous mixture of gas and dust from the galaxy due to SNII, (b) the dust distribution radius, $r_d$, is taken to scale with the effective stellar distribution scale, $r_e$, such that $r_d \approx (0.6,1.0) r_e$ at $z \approx (5.7,6.6)$ respectively; the calculation of $r_e$ is described later in Sec. \ref{model}, (c) $f_c$ is calculated assuming a slab-like dust distribution and we use $f_\alpha = (1.5,0.6) f_c$, as inferred for LAEs at $z \approx (5.7,6.6)$, and (d) $T_\alpha$ is calculated using the mean photoionization rate predicted by the Early Reionization Model (ERM, reionization ends at $z \approx 7$) as proposed by \citet{gallerani2008}, according to which the neutral hydrogen fraction $\chi_{HI} = (6.0 \times 10^{-5}, 2.3 \times 10^{-4})$ at $z \approx (5.7, 6.6)$. 

Once the above calculations have been carried out, following the current observational criterion, galaxies with $L_\alpha \geq 10^{42.2} {\rm erg \, s^{-1}}$ and an observed Ly$\alpha$ equivalent width $EW = L_\alpha/L_c \geq 20$ \AA\, are identified as LAEs. Complete details of these calculations can be found in \citet{dayal2010a} and \citet{dayal2011}.

\section{Molecular hydrogen physics}
\label{h2_physics}

$\rm{H_2}$ can be formed in galaxies by two main methods: the first, and rather inefficient method involves gas-phase reactions mainly through the coupled reactions $\rm{e^{-}+H\rightarrow H^{-} + h\nu}$ and $\rm{H^{-}+H\rightarrow H_{2} + e^{-}}$ \citep{mcdowell1961,palla1983}. The second, more 
efficient channel is through bond formation on dust grains: this process begins with the collision and absorption of at least two hydrogen atoms by the same dust grain. The hydrogen atoms are weakly bound to the grain surface through Van der Waals forces and can migrate on the grain either by tunnelling or thermal hopping. If the hydrogen atoms encounter each other, bond formation takes place, the excess energy is released to the grain, and the $\rm{H_2}$ molecule is ejected into the gas phase \citep{gould1963}. However, $\rm{H_2}$ molecules so produced can be dissociated by the far ultraviolet (FUV) interstellar radiation field in the Lyman-Werner (LW) band between 11.2-13.6 eV; the twin processes of self shielding and dust absorption (e.g. \citealt{hollenbach1999}) drive the shielding of $\rm {H_2}$ to FUV photons, thereby preventing photodissociation. Considering these processes is of utmost importance since $\rm{H_2}$ is found in molecular clouds that are surrounded by a photodissociation region (PDR) where the gas is predominantly atomic. 

In this work we estimate the $\rm{H_2}$ mass of each LAE at $z \approx (5.7,\,6.6)$ using the analytic model presented in \citet{krumholz2008, krumholz2009} and \citet{mckee2010}, hereafter referred to as the KMT model. In brief, the KMT model considers an idealized spherical cloud immersed in a uniform, isotropic LW radiation field. Then, the equations of radiative transfer coupled to the $\rm{H_2}$ formation-dissociation balance are solved, assuming the cloud to be in the steady state. 

The analytical solution to the $\rm{H_2}$ mass fraction, $f_{\rm{H_2}}=M_{\rm{H_2}}/M_{\rm{HI}}$, is then obtained by solving for the radial position at which the transition between the atomic envelope and the molecular interior occurs within the cloud; in this equation $M_{\rm {H_2}}$ and $M_{\rm{HI}}$ refer to the mass of molecular hydrogen and the mass of neutral hydrogen respectively. The KMT study shows that the fraction of the radius at which this transition occurs is solely a function of the dust optical depth in the LW band and the dimensionless parameter $\chi$, which are discussed in what follows.

\subsection{Modelling molecular hydrogen in LAEs}
\label{model}
We now describe the model used to calculate the $\rm{H_2}$ content of the LAEs identified in the simulation snapshots at $z \approx 5.7, 6.6$. We start by assuming that the MCs lie in a region that extends from the centre of the galaxy up to the effective stellar distribution radius, $r_e$, calculated in \citet{dayal2010a}. This assumption has been motivated by the fact that star formation occurs in MCs; the physical distribution scale of MCs and the stars is therefore expected to be quite similar. Further, the value of $r_e$ is based on estimates following the results of \citet{bolton2008}, who have derived fitting formulae relating the V-band luminosity and the stellar distribution scale from their observations of massive, early type galaxies between $z=0.06-0.36$. Though not an entirely robust estimate, we extend this result to galaxies at $z \approx 5.7$ and $6.6$ due to the paucity of observational data regarding the stellar distribution scales in high-redshift galaxies. However, such estimates are in surprisingly good agreement with recent observational results: \citet{malhotra2011} find that the half-light radius of LAEs has a mean value $\approx 0.16$ arcsec at $z \approx 5.7$ and this remains constant for all redshifts in the range $2 \leq z \leq 6.5$; these estimates lie within 1$\sigma$ of the mean value of the theoretical $r_e$ estimates used throughout this paper.

As mentioned above, using the KMT model, the $\rm{H_2}$ mass fraction solely depends on the dust optical depth in the LW band, $\tau_c$, and the dimensionless parameter $\chi$, such that the analytical solution for the $\rm{H_2}$ fraction can be written as \citep{mckee2010}:
\begin{equation}
\label{fH2}
f_{\mathrm{H_2}}\simeq 1- \left(\frac{3}{4}\right) \frac{s}{1+0.25s},
\end{equation}
where the dimensionless parameter $s$ can be expressed as
\begin{equation}
\label{s}
s=\frac{\ln(1+0.6\chi+0.01\chi^2)}{0.6\tau_c}.
\end{equation}
Here, 
\begin{equation}
\label{chi}
\chi = \frac{f_{diss} E_0 \sigma _d c} {n_{\rm{CNM}} \mathcal{R}}, 
\end{equation}
where, $f_{diss}\simeq 0.1$ \citep{draine1996} is the fraction of absorbed LW band photons that lead to dissociation of the $\rm{H_2}$ molecules, $E_0$ is the free space photon number density in the LW band, $\sigma_d$ is the dust absorption cross-section per hydrogen nucleus to LW photons, $c$ is the speed of light, $n_{\rm{CNM}}$ is the number density of gas in the cold atomic medium that surrounds the molecular part of the cloud, and $\mathcal{R}$ is the coefficient for the 
rate of $\rm{H_2}$ formation on the surface of dust grains. 

It must be noted that these equations apply only for $s<2$; for values of $s\geq2$, the gas is predominantly atomic, such that $f_{\rm{H_2}}=0$. Also, we note that the calculations presented here concern only average quantities in a spherically symmetric framework; a full calculation of the radial dependence of the model parameters is the subject of ongoing work. We now explain the calculations of $E_0$, $\sigma_d$, $n_{\rm{CNM}}$, $\mathcal{R}$ and $\tau_c$ in what follows.

\subsubsection{LW photon number density}
As mentioned in Sec. \ref{identify_laes}, we obtain the intrinsic spectrum of each LAE using the population synthesis code {\tt STARBURST99}. Then, assuming all the stars to form at the centre of the galaxy, the number density of LW photons of a specific wavelength $\lambda$ ($912$ \AA\, $\leq \lambda \leq 1120$ \AA), at a distance $r$ from the centre can be expressed as
\begin{equation}\label{number}
n_\lambda(r)=\frac{L_{\lambda}}{4  \pi c r^2 }\left(\frac {\lambda}{hc}\right).
\end{equation}
Here, $L_\lambda$ is the monochromatic luminosity at the wavelength $\lambda$ obtained using {\tt STARBURST99} and $h$ is the Planck constant. The free space photon number density at radius $r$, in the entire LW band, $n_{LW}$, can then be calculated by integrating over all the wavelengths in the band such that
\begin{equation}
n_{LW}(r)= \frac{1}{4 \pi r^2 c}\int_{912 {\AA}}^{1120{\AA}} \frac{\lambda L_\lambda}{hc } d \lambda.
\end{equation}
The value of $n_{LW}$ averaged over a sphere of radius $r_e$ then gives the photon number density in the LW band such that
\begin{equation}
E_0 \equiv \langle n_{LW} \rangle =3n_{LW}(r_e) 
\end{equation}

\subsubsection{Cold neutral medium density}
\begin{table} 
\begin{center} 
\caption {As a function of the halo mass range (col. 1), we show the fraction of ISM gas with temperature $T \leq 10^4\,\rm{K}$ (col. 2).}
\begin{tabular}{|c|c|c|c} 
\hline 
$M_h$ & $f_4$  \\
$[M_\odot]$ & $ $ \\ 
\hline
$ <10^{10}$ & $0.58$\\
$10^{10-10.4}$ & $0.40$\\ 
$10^{10.4-10.8}$ & $0.35$\\ 
$10^{10.8-11.2}$ & $0.35$\\ 
$>10^{11.2}$ & $0.35$\\
\hline
\label{table1} 
\end{tabular} 
\end{center}
\end{table} 

Much of the neutral gas in galaxies is observed to be cold, with temperatures of order of $\sim 100\,\rm{K}$ (CNM), or warm, with temperatures of order of $\sim 10^4\,\rm{K}$ (WNM), in approximate pressure balance \citep{wolfire2003}; as has been pointed out before, MCs form in regions where the gas is primarily cold. To calculate the density of the cold neutral medium, we start by obtaining the fraction, $f_4$, of ISM gas with temperature $T \leq 10^4\,\rm{K}$ \citep[see also Fig. 1, ][]{dayal2010a}; the value of $f_4$ averaged over galaxies of different halo masses is shown in Tab. \ref{table1}. However, the large volume simulated ($\approx 10^6 \, \rm{cMpc^3}$), naturally results in a low mass resolution, such that we are unable to resolve the cold and warm gas phases inside the ISM of individual galaxies. We therefore make the approximation that the ISM of each simulated galaxy has an equal amount of cold and warm neutral gas. The mass of the cold neutral gas, $M_{\rm{CNM}}$, in any galaxy can then be calculated as

\begin{equation}
M_{\rm{CNM}}=\frac{1}{2} f_4 M_{\mathrm{H}},
\end{equation}
where $M_{\rm{H}}=0.76\,M_{gas}$ is the mass of hydrogen in the ISM. We assume the gas to be distributed in a disk with a radius $r_g$ and scale height $H$ such that \citep{ferrara2000}

\begin{eqnarray}\label{rad_ht}
r_g & = & 4.5 \lambda r_{200}, \\
\frac{H}{r_g} & = & 15.3\lambda \left(\frac{c_s}{v_e}\right)^2
\end{eqnarray}

Here, value of the spin parameter is taken to be $\lambda=0.04$ \citep{ferrara2000} and the virial radius, $r_{200}$, is calculated assuming the collapsed region has an overdensity of 200 times the critical density at the redshift considered. Assuming that the typical velocity that determines the scale height of the disk is that of WNM, the effective gas sound speed is taken to be $c_s=10\, \rm{km\, s^{-1} }$. Finally, the halo escape velocity, $v_e$, is related to the circular velocity of the halo, $v_c$, by the relation $v_e=2pv_c$ with $p=1.65$ \citep{maclow1999}. 
The average global number density of the CNM, $n_{\rm{CNM}}$ can then be expressed as
\begin{equation}
n_{\rm{CNM}}=\frac{M_{\rm{CNM}}}{\pi r_g^2 H m_H},
\end{equation}
where $m_H$ is the hydrogen mass.
\subsubsection{Molecular hydrogen formation rate}
The rate of $\rm{H_2}$ formation on dust grains, $\mathcal{R}$, can be expressed as \citep{hirashita2005}
\begin{equation}\label{rate}
\begin{split}
\mathcal{R}=4.1 \times 10^{-17} S \left(\frac{a}{0.1 \, \mu \mathrm{m}}\right)^{-1}\left(\frac{\mathcal{D}}{10^{-2}}\right)\\
 \times \left(\frac{T}{100 \, \mathrm{K}}\right)^{1/2}\left(\frac{\delta}{2 \, \mathrm{g cm^{-3}}}\right)^{-1}\, \mathrm{cm^3 s^{-1}},
\end{split}
\end{equation}
where $S=S(T,T_d)$ is the sticking coefficient for hydrogen atoms on dust grains, $T$ is the gas temperature, $\delta$ is the density of the dust grains, $\mathcal{D}$ is the dust-to-gas ratio, and $a$ is the radius of the dust grain which is assumed to be a sphere. Following the assumptions of \citet{dayal2010a}, we assume all the dust grains to be carbonaceous such that $\delta = 2\, \rm{g\,cm^{-3}}$; using the size distribution of SNII-produced dust grains \citep{todini2001}, we use an average grain size value of $a  = 300\,\rm{\AA}$. As mentioned before, dust predominantly forms in high-density, cold MCs which have a more effective self-shielding to LW photons compared to the lower density WMN; following this argument, we use $T = 100\,\rm{K}$. 

Further, the sticking coefficient $S(T, T_d)$ is given by \citet{hollenbach1979} and \citet{omukai2000} as
\begin{equation}\label{sticking2}
\begin{split}
S(T, T_d) = [1 + 0.04(T + T_d )^{0.5} + 2 \times 10^{-3}T + 8 \times 10^{-6}T^2]^{-1}\\
\times \{1 + exp[7.5 \times 10^2(1/75 - 1/T_d)]\}^{-1},
\end{split}
\end{equation}
where $T_d$ is the dust temperature. However, the sticking co-efficient is not affected by the exact dust temperature as long as $T_d \leq 70\,\rm{K}$, which is true for all LAEs, which have dust temperatures $T_d \approx 30-45\,\rm{K}$. 

\subsubsection{Dust cross-section and optical depth}
\label{dust}

We now discuss the calculation of the dust cross-section per hydrogen atom to LW photons, which can be expressed as 
\begin{equation}
\label{sigmad}
\sigma_d = \frac{Q_{abs}(a) \pi a ^2}{N_H},
\end{equation}
where $a  = 300\,$\AA\, is the average radius of SNII produced dust grains as mentioned above, $Q_{abs}(a)$ is the effective cross-section normalised to the geometric one for the average grain size and a wavelength of $1000\,$\AA\, corresponding to the centre of the LW band \citep{draine1984} and $N_{H}$ is the number of hydrogen nuclei per dust grain. To calculate $N_H$, we use the dust to gas ratio, $\mathcal{D}$ such that
\begin{equation}\label{absorption}
\mathcal{D}=\frac{M_{dust}}{M_{gas}}\simeq \frac{\frac{4}{3}\pi a^3 \delta N_{dust}}{\mu m_p N_H}
\end{equation}
where $\mu=0.59$ is the mean molecular weight of a fully ionized gas of primordial composition. Substituting $N_H$ from Eq. \ref{sigmad} into Eq. \ref{absorption} yields
\begin{equation}\label{sigmapolvere}
\sigma _ d= \frac{3}{4} \frac{\mu m_p \mathcal{D}}{\delta} \frac{Q_{abs} (a)}{a}.
\end{equation}
From Eqns. \ref{rate} and \ref{sigmapolvere}, we note that $\chi$ is independent of $\mathcal{D}$.

Finally, the dust optical depth in the LW band is calculated as:
\begin{equation}
\label{tauc}
\tau _{c} = \frac{\Sigma _d Q_{abs} (a)}{\frac{4}{3} a \delta},
\end{equation}
where the dust surface density $\Sigma_{d}=M_{d}/(\pi r_{d}^2$) and $r_d \approx (0.6, 1.0) \times r_e$ at $z\approx(5.7, 6.6)$ as required to best fit the LAE 
UV luminosity functions \citep{dayal2010a}.

Once these calculations have been carried out, the total molecular hydrogen mass, $M_{\rm{H_2}}$ in each LAE is estimated as
\begin{equation}\label{massh2}
M_{\rm{H_2}}=f_{\rm{H_2}} M_{\rm{HI}},
\end{equation} 
where $M_{\rm{HI}} = f_4 M_{H}$ is the neutral hydrogen mass in the gas disk within a radius $r_e$ and scale height $H$, where star formation takes place. 

\section{Results}
Once that the above calculations have been carried out, we can discuss the results regarding the molecular hydrogen fraction and the total $\rm{H_2}$ mass, and relate these to the physical properties of LAEs at $z \approx 5.7, 6.6$. In what follows, we also calculate the visibility of such LAEs through their CO emission and end by making predictions for such detections in the ALMA Early Science Release (ESR). 

\begin{table*}
\begin{center} 
\caption {For all the LAEs at the redshifts shown (col. 1), we show the range of stellar mass (col. 2), the range of SFR (col. 3), the range of cold neutral medium mass (col. 4), the range of dust mass (col. 5), the average molecular fraction (col. 6), the average mass of molecular hydrogen (col. 7), the average intrinsic CO(1-0) luminosity (col. 8), and the average value of CO(6-5) flux (col. 9).
} 
\label{table2}
\begin{tabular}{l*{7}{c}r}
\hline
$z$ & $M_{*}$ & $\dot{M_{*}}$ & $M_{\rm{CNM}}$ & $M_d$  & $\langle f_{\rm{H_2}} \rangle$ & $\langle M_{\rm{H_2}}\rangle$ & $\langle L_{\rm{CO}} \rangle$ & $\langle S_{\rm{CO}} \rangle$\\
$ $	& $[\rm{M_{\odot}}]$ & $[\rm{M_{\odot}\,yr^{-1}}]$ & $[\rm{M_{\odot}}]$ & $[\rm{M_{\odot}}]$  &$ $& $[\rm{M_{\odot}}]$ & $[\rm{K\,\,km/s\,\,pc^2}]$ & $[\rm{mJy}]$ \\
\hline  
$5.7$ & $10^{[8.0 - 10.4]}$  & $0.8-120$ & $10^{[8.4-10.1]}$ & $10^{[3.4-7.2]} $ & $0.6$ & $10^{8.9}$  & $10^{9.0}$  & $0.2$  \\           
$6.6$ & $10^{[8.1 - 10.0]}$ & $1.6-46.4$ & $10^{[8.5-9.8]}$ & $10^{[4.0-6.9]}$  & $0.3$ &  $10^{8.4}$ & $10^{8.5}$ & $0.06$     \\
\hline
\end{tabular}
\end{center}
\end{table*}

\subsection{Molecular hydrogen content of LAEs}
\label{gal_prop}
As shown in Sec. \ref{model}, the ${\rm H_2}$ fraction is decided by three important physical parameters: (a) the SFR which determines the intensity of the ${\rm H_2}$-dissociating LW field, (b) the amount of cold gas available to shield the MC against the LW field, and (c) the dust mass on which the ${\rm H_2}$ forms, and which additionally shields the molecular hydrogen by absorbing LW photons. We now quantify how the molecular hydrogen fraction depends on each of these parameters.

\begin{figure}
\begin{center}
\includegraphics[scale=0.42]{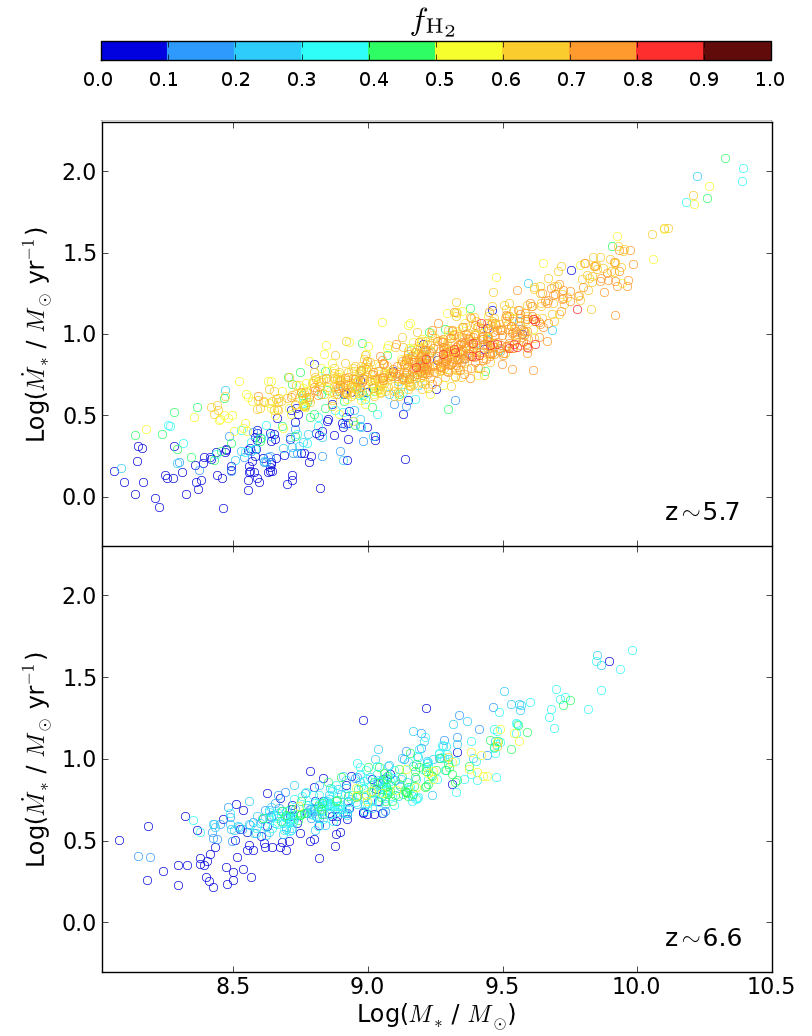}
\end{center}
\captionsetup{singlelinecheck=off}
\caption[]{\footnotesize The molecular hydrogen fraction, $f_ {\rm{H_2}}$, of LAEs at $z \approx 5.7$ (6.6) as a function of the stellar mass ($M_*$) and SFR ($\dot M_*$) are shown in the upper (lower) panels respectively. Points are color-coded for different values of $f_{\rm{H_2}}$.}\label{fh2_sfr_mstar}
\end{figure}

We start our discussion by mentioning that the stellar mass of LAEs ranges between $M_* = 10^{8.0-10.4} M_\odot$ at $z \approx 5.7$. As expected in a hierarchical structure formation scenario where progressively larger objects form from the merger of smaller ones, such range narrows to $M_* \approx 10^{8.1-10} M_\odot$ at $z \approx 6.6$ (see also Tab. \ref{table2}); galaxies with stellar masses above $10^{10} M_\odot$ have not had the time to assemble in large numbers by $z \approx 6.6$. Further, the SFR of LAEs falls in the interval $\dot M_* \approx 0.8-120 M_\odot \, {\rm yr^{-1}}$ at $z \approx 5.7$, with larger ($M_* \geq 10^{9.5} M_\odot$) galaxies being the most efficient in star formation. At stellar masses lower than this value, the SFR-stellar mass relation flattens at both the redshifts considered, as seen from Fig. \ref{fh2_sfr_mstar} due to the stronger effects of mechanical feedback (ejection of gas in outflows) inhibiting star formation \citep[see][]{dayal2009}. As a result of the narrower stellar mass range, the SFR for $z \approx 6.6$ are concentrated in a narrower range between $1.6-46\,  {\rm M_\odot \, yr^{-1}}$ (see Tab. \ref{table2}).

As for the molecular hydrogen fraction, starting with $z \approx 5.7$, we find $f_{\rm H_2}\leq 0.1$ for galaxies with $M_* < 10^9 M_\odot$; it rises to $\approx 0.5-0.9$ for intermediate mass galaxies with $M_* = 10^{9-10} M_\odot$ and then decreases again to 0.2-0.6 for the few largest galaxies. Such behavior can be explained as follows: although the smallest galaxies ($M_*<10^9 M_\odot$) have the smallest SFR (and hence weakest LW field), they also less dusty, resulting 
in a lower ${\rm H_2}$ production rate and self-shielding ability against photodissociation. On the other hand, the larger SFR compared to the dust and cold gas mass in the largest galaxies leads to a decreased $f_{\rm H_2}$. It is thus the intermediate mass galaxies that show the largest ${\rm H_2}$ fraction by virtue of a well-tuned balance between the ${\rm H_2}$ formation and dissociation rates. Such argument is supported by the fact that at a given value of the stellar mass, intermediate mass galaxies with the lowest SFR have the largest ${\rm H_2}$ fraction as seen from Fig. \ref{fh2_sfr_mstar}. Qualitatively, the situation remains the same at $z \approx 6.6$, although quantitatively, the $f_{\rm H_2}$ value is lower for all LAEs; the reason for this is detailed in what follows.

\begin{figure}
    \begin{center}
\includegraphics[scale=0.42]{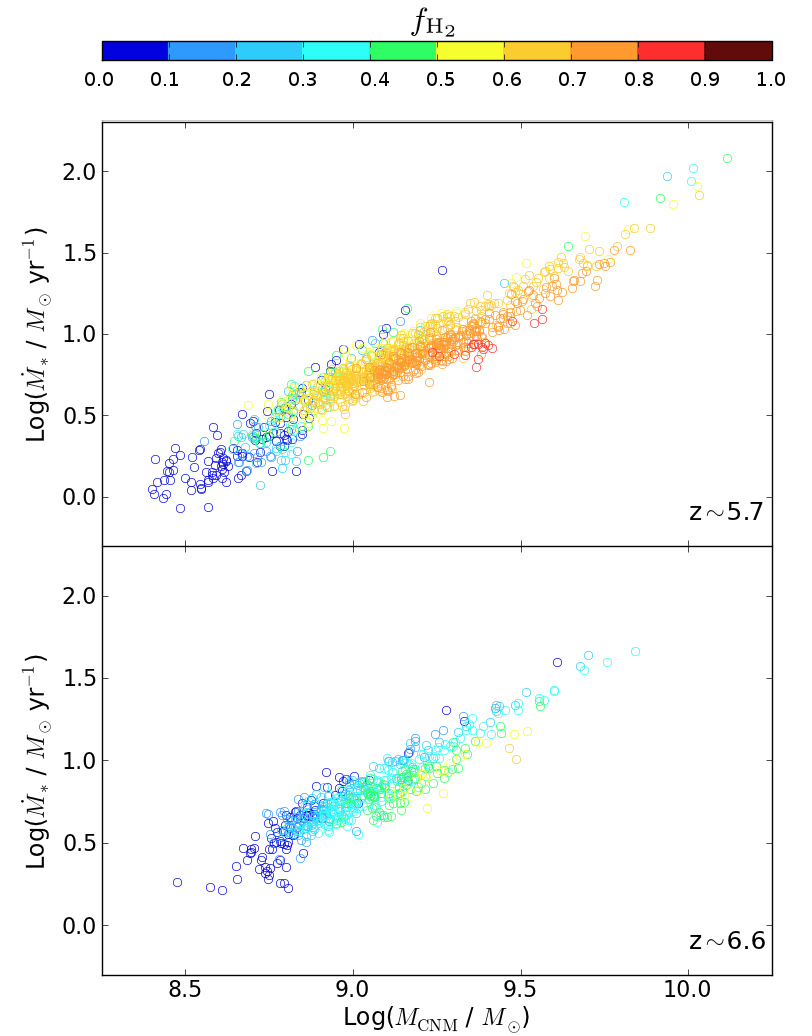}
	\end{center}
	\captionsetup{singlelinecheck=off}
    \caption[]{\footnotesize The molecular hydrogen fraction, $f_ {\rm{H_2}}$, of LAEs at $z \approx 5.7$ (6.6) as a function of the stellar mass ($M_*$) and mass of cold \HI ISM gas with temperature less than 100 K ($M_{\rm CNM}$) are shown in the upper (lower) panels respectively. Points are color-coded for different values of $f_ {\rm{H_2}}$.} \label{sfr_mcold_fh2}
\end{figure}

To understand the $f_{\rm H_2}$ dependence on the cold gas mass ($M_{\rm{CNM}}$), we start by noting that $M_{\rm{CNM}}$ scales well with the SFR as shown in Fig. \ref{sfr_mcold_fh2}; more massive galaxies have a smaller cold mass fraction, possibly due to stellar sources heating a greater part of the ISM to higher temperatures (see also Tab. \ref{table1}). From the same figure, we see that for a given value of $M_{\rm{CNM}}$, galaxies with the lowest SFR have the largest value of $f_{\rm H_2}$ which is easy to understand considering that for a given $M_{\rm{CNM}}$, at lower SFR, the ${\rm H_2}$ LW dissociation becomes less efficient. Again, it is the intermediate mass galaxies that have the largest value of $f_{\rm H_2}$, while the ${\rm H_2}$ fraction is larger at $z \approx 5.7$ compared to $z \approx 6.6$ for all LAEs.

\begin{figure}
   \begin{center}
 \includegraphics[scale=0.42]{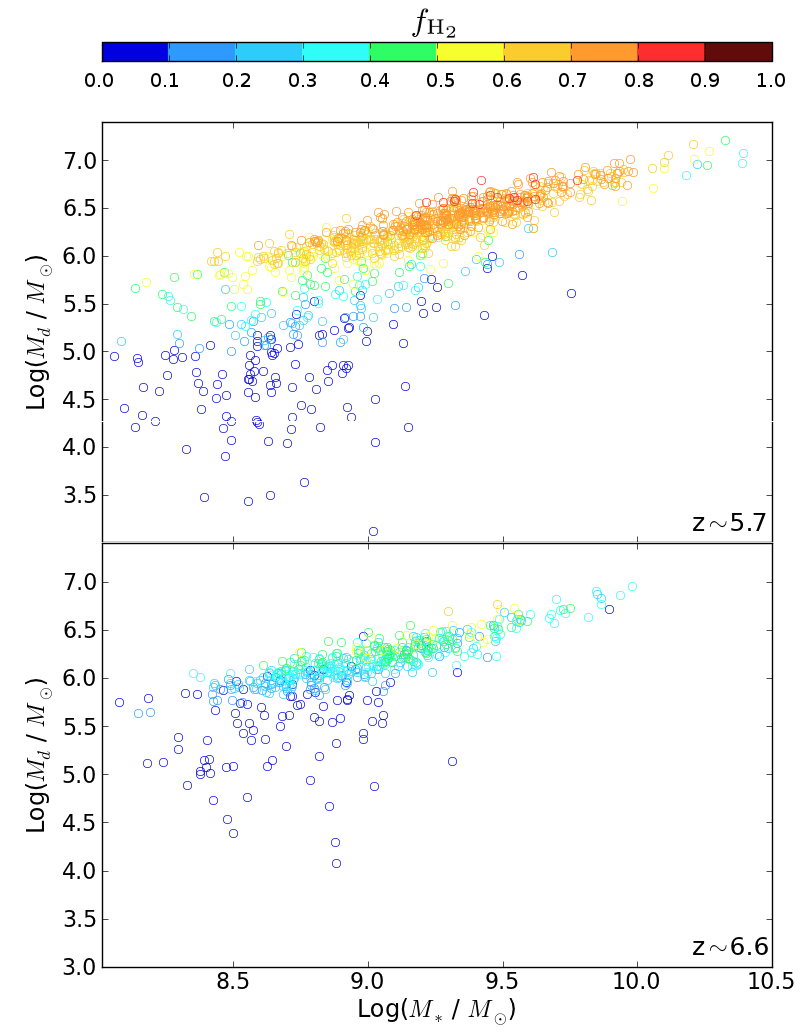}  
	\end{center}
	\captionsetup{singlelinecheck=off}
    \caption[]{\footnotesize The molecular hydrogen fraction, $f_ {\rm{H_2}}$, of LAEs at $z \approx 5.7$ (6.6) as a function of the stellar mass ($M_*$) and dust mass ($M_d$) are shown in the upper (lower) panels respectively. Points are color-coded for different values of $f_ {\rm{H_2}}$.}\label{mdust_mstar_fh2}
\end{figure}

Dust also plays a key role in terms of the ${\rm H_2}$ abundance since this molecule predominantly forms on dust grains which also shield the molecular hydrogen so formed, by absorbing LW photons. The dust mass for each LAE has been calculated as explained in Sec. \ref{identify_laes}; since we assume stellar sources (i.e. SNII) to be the main dust producers, the total dust mass scales well with the stellar mass, ranging between $10^{3.4-7.2} M_\odot$ at $z \approx 5.7$. As mentioned before, since ${\rm H_2}$ forms on dust grains, naively it might be expected that the larger the dust content, the larger the value of $f_{\rm H_2}$. However, this does not hold true; the LW intensity in the largest galaxies is enough to dissociate the ${\rm H_2}$ formed as a result of which, the intermediate mass galaxies end up with the largest ${\rm H_2}$ fraction. As expected, at a given value of $M_*$, galaxies with the largest dust mass have the largest $f_{\rm H_2}$ value (Fig. \ref{mdust_mstar_fh2}).

\begin{figure}
\centering
\includegraphics[scale=0.4]{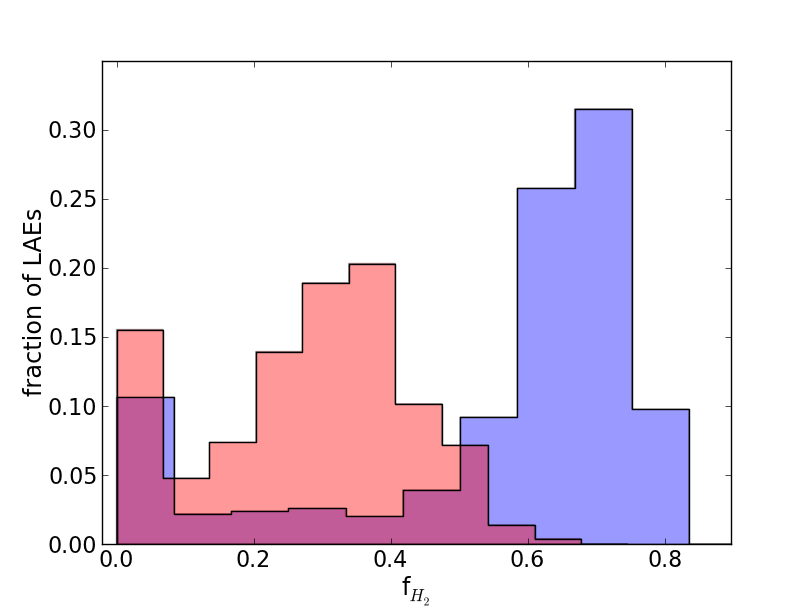}
\captionsetup{singlelinecheck=off}
\caption[]{Normalized distribution of the number of LAEs as a function of the molecular hydrogen fraction $f_{\mathrm{H_2}}$ at $z\approx5.7$ (dark/blue) and $z\approx6.6$ (light/red).} \label{molecular_fraction}
\end{figure}

Finally, we discuss why the value of $f_{\rm H_2}$ is lower at $z \approx 6.6$ compared to $z \approx 5.7$, even though galaxies identified as LAEs are extremely similar at these redshifts, in terms of $M_*, \dot M_*, M_d$ and $M_{\rm{CNM}}$ (see Tab. \ref{table2}): as mentioned in Sec. \ref{dust}, the value of the dust distribution radius, $r_d = (0.6,1.0) r_e$ at $z \approx (5.7, 6.6)$, is fixed by matching the observed LAE UV luminosity functions. These values imply that the dust is more concentrated in the inner parts of LAEs than the stars themselves, perhaps hinting at the existence of dust/metallicity radial gradients. We note that $r_e$, the stellar distribution scale, is similar at both the redshifts considered. The  larger value of $r_d$ results in a dust optical depth that is about 1.6 times smaller at $z \approx 6.6$ as compared to that $5.7$ (see Eq. \ref{tauc}); although comparable ${\rm H_2}$ masses would be produced in LAEs with similar physical properties at both $z \approx 5.7$ and $6.6$, a larger amount is dissociated at $z \approx 6.6$ due to a decreased dust absorption of ${\rm H_2}$-dissociating LW photons. Averaged over all LAEs, $f_{\rm H_2} \approx 0.6$ at $z \approx 5.7$, and only $f_{\rm {H_2}} \approx 0.3$ at $z \approx 6.6$, as seen from Fig. \ref{molecular_fraction} and Tab. \ref{table2}. From the same figure, we see that while $f_{\rm H_2}$ covers the broad range $0-0.85$ at $z \approx 5.7$, no LAEs have $f_{\rm H_2} \geq 0.65$ at $z \approx 6.6$, as a result of the smaller dust optical depth. Finally, translating the value of $f_{\rm H_2}$ into a total \mhh\, (Eq. \ref{massh2}), we find the average value of $M_{\rm H_2} \approx 10^{8.9}, 10^{8.4}$ at $z \approx 5.7$ and 6.6 respectively, as shown in Tab. \ref{table2}.

To summarize, we find that intermediate mass LAEs ($M_* \approx 10^{9-10} M_\odot$) have the largest molecular hydrogen fraction; a delicate balance between $\dot M_*, M_{\rm{CNM}}$ and $M_d$ leads to ${\rm H_2}$ formation (and self-shielding) dominating over ${\rm H_2}$ photo-dissociation. Finally, due to a smaller dust optical depth (by a factor $\approx 1.6$) at $z \approx 6.6$, the ${\rm H_2}$ fraction is about twice as large at $z \approx 5.7$ for LAEs with comparable physical properties.

\subsection{CO detectability in LAEs}
\label{co_detectability}

Now that the dependence of $f_{\rm H_2}$ on the physical properties of LAEs has been understood, we are in a position to make predictions for the ${\rm H_2}$ detectability in these galaxies. ${\rm H_2}$  has strongly forbidden rotational transitions; the rotational-vibrational lines have very high excitation temperatures, that are attainable only under somewhat extreme conditions involving intense irradiation or shock waves \citep[see][]{solomon2005} in the absence of which the ${\rm H_2}$ is invisible. On the other hand, CO has a weak dipole moment: its rotational levels are then easily excited and thermalized by collisions with $\rm{H_2}$. In addition, CO is a very stable molecule and the most abundant after ${\rm H_2}$ \citep[e.g.][]{solomon2005,omont2007}. Because of such considerations, it is popularly used as a tracer of ${\rm H_2}$. 

\begin{figure}
 \includegraphics[scale=0.42]{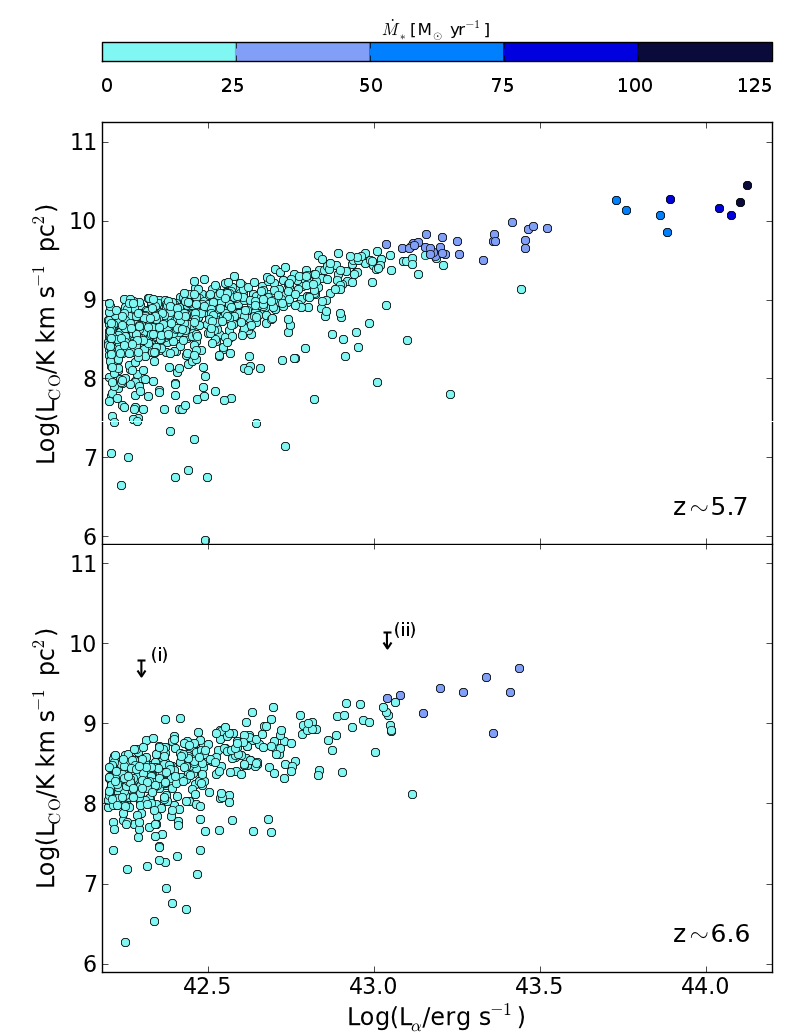}
 \captionsetup{singlelinecheck=off}
  \caption[]{\footnotesize As a function of the observed Ly$\alpha$ luminosity ($L_\alpha$), we show the estimated CO luminosity at $z \approx 5.7$ and 6.6 in the upper and lower panels respectively; the points shown are color-coded according to the SFR. In the lower panel, the arrows indicate the upper limit on the $L_{\rm{\rm{CO}}}$ obtained by \citet{wagg2009}: $(i)$ corresponds to the nondetection of the CO (1-0) line of HCM 6A, a LAE at $z\approx6.6$, with $\rm{SFR}\approx2\,\,\rm{M_{\odot}yr^{-1}}$ \citep{hu2002}. $(ii)$ corresponds to the nondetection of the CO (1-0) line of IOK-1, a $z\approx7$ LAE with $\rm{SFR}\approx10\,\,\rm{M_{\odot}yr^{-1}}$ \citep{iye2006}.}
  \label{lcosfr}
\end{figure}

The luminosity of the CO(1-0) transition can be related to \mhh\, mass as
\begin{equation}
L_{\rm{CO}}=M_{\rm{H_2}}/\alpha.
\end{equation}
The parameter $\alpha$ used in the equation above depends on the distribution of star forming clouds: in the Milky Way, where star formation takes place in molecular clouds with dense cores, confined by self gravity, $\alpha=4.6\,\rm{M_{\odot}\, K \, km \, s^{-1} \, pc^2}$ \citep{solomon1987}. 
On the other hand, in high-redshift Ultra Luminous Infrared Galaxies (ULIRGs), where star formation is expected to occur in a dense intercloud medium bound by the potential of the galaxy, $\alpha$ has a much lower value of $0.8\,\rm{M_{\odot}\, K \, km \, s^{-1} \, pc^2}$ \citep{downes1998}. It is worth noting that the value of $\alpha$ is also related to the metallicity of the interstellar gas
\citep[e.g.][]{leroy2009, narayanan2011, genzel2011}. Using the Green Bank telescope, \cite{wagg2009} undertook a search for CO emission in two LAEs at $z >6$, and they adopted the ULIRG value of the conversion factor to estimate \mhh. For a reasonable comparison with such LAE data, we use the same value of $\alpha=0.8\,\rm{M_{\odot}\, K \, km \, s^{-1} \, pc^2}$ to compute the CO(1-0) luminosity values for simulated LAEs at $z \approx (5.7,\, 6.6)$. We find that the value of $L_{\rm{CO}}$ scales with $L_\alpha$ for both the redshifts considered as shown in Fig. \ref{lcosfr}; quantitatively we find, $L_{\rm{CO}} \propto L_{\alpha}^{(1.04,\,1.07)}$ at $z\approx (5.7,\, 6.6)$ respectively. Although such a relation has a huge scatter, this implies that the brightest LAEs are the best candidates for molecular emission searches.

Interestingly, such a trend also implies that although the value of $f_{\rm{H_2}}$ is the largest for the intermediate mass/luminosity LAEs (see Sec. \ref{gal_prop}), the increasingly large neutral hydrogen mass, $M_{\rm{HI}}$, in the star forming disk (see Eq. \ref{massh2}) of massive galaxies wipes out such a subtle signature, whose imprint remains as a flattening of the $L_{\rm{CO}}- L_\alpha$ relation towards the most luminous objects. As expected, as a result of their lower dust optical depth, and hence, a lower molecular hydrogen fraction, the CO luminosity for LAEs is smaller at $z \approx 6.6$ than at $z \approx 5.7$; averaged over all LAEs, the CO luminosity at $z \approx 6.6$ is about a factor 3 lower than that at $z \approx 5.7$, as shown in Tab. \ref{table2}. As a validation of our model, our theoretical $L_{\rm{CO}}$ estimates are in accord with the upper limits found by \cite{wagg2009} for their observed LAEs, as seen from lower panel of Fig. \ref{lcosfr}, which represents an encouraging sanity check of our model. Finally, we note that $\alpha$ is expected to be larger than the Galactic value in low metallicity environments \citep{leroy2009, narayanan2011, genzel2011}, which is likely the case for LAEs, while we have used a value derived using ULIRGs. As expected, an increase in the value of $\alpha$ would lead to a decrease in the CO luminosity, thereby negatively affecting the CO detectability of LAEs. 

\subsection{Predictions for ALMA early science}
\label{ers}
 
Now that the CO(1-0) luminosity has been calculated for all LAEs in our simulation, we make predictions for the detectability of such CO lines using ALMA. As of now, these observations have been limited to the most luminous high-$z$ sources such as QSOs \citep[e.g.][]{cox2002, bertoldi2003, walter2004, weiss2007, wang2010, riechers2011b} and sub-millimeter galaxies (SMG) \citep[e.g.][]{greve2005, tacconi2006}; normal galaxies such as Lyman break galaxies (LBGs) have only been detected at a much lower redshifts, $z\approx3$. We start by noting that ALMA Early Science (Cycle 0) consists of the use of 16 antennas and a limited number of Bands. In this configuration, the lowest CO rotational transition observable at $z \approx 5.7$ is CO(5-4) ($\nu_{rest}=576.267\,\rm{GHz}$), which falls within ALMA Band 3 ($84 - 116\,\rm{GHz}$). However, as is shown in what follows, the rotational transition strength of the CO(5-4) line is quite comparable to that of CO (6-5). The latter is the lowest line observable with ALMA Cycle 0 for both ($z \approx 5.7,\,6.6$) samples, for which reason we show results for the CO(6-5) transition in what follows; in the future, using the the full capabilities of ALMA, lower frequency bands ($31.3-45,\rm{GHz}$ and $67 - 90\,\rm{GHz}$) will also allow for the study of lower-J CO line transitions at high redshift.\\
We now describe how we translate the CO(1-0) luminosities calculated above in Sec. \ref{co_detectability} in to CO(6-5) luminosities.
Theoretical fits to observational CO spectral energy distributions (SEDs) have been carried out by several authors using Local Thermal Equilibrium \citep[LTE; ][]{obreschkow2009} or Large Velocity Gradient (LVG) models \citep{bayet2009} to describe the molecular gas. We use the model proposed by \citet{obreschkow2009} which assumes a single gas component in LTE. In this framework, the frequency integrated luminosity from the transition ($J \rightarrow J-1$) can be expressed as \citep{obreschkow2009}:
\begin{equation}\label{lj}
L(J \rightarrow J-1) \propto 1 - \mathrm{exp} (\tau_J) \cdot \frac{J^4}{\mathrm{exp}^{(\frac{h\nu_{\rm{CO}} J}{k_b T_e})}-1},
\end{equation}
where $T_e=100\,\rm{K}$ is the gas excitation temperature, $\nu_{\rm{CO}}=115.271\,\rm{GHz}$ is the rest-frame frequency of the CO(1-0) transition, $\tau_J$ is the optical depth and $k_b$ is the Boltzmann constant. Further, $\tau_J$ can be expressed as 
\begin{equation}\label{tauj}
\tau_J = 7.2 \overline{\tau} \mathrm{exp}(-\frac{h\nu_{\rm{CO}} J^2}{2 k_b T_e}) \mathrm{sinh}\bigg(\frac{h\nu_{\rm{CO}} J}{2 k_b T_e}\bigg),
\end{equation}
where $\overline{\tau}$ is an experimental determined normalization constant, which we take to be $\overline{\tau}=2$, following the results of \citet{obreschkow2009}. 

Considering that the frequency of the ($J \rightarrow J-1$) transition is related to that of (1-0) by $\nu_J= J \nu_{\mathrm{CO}}$, and using the relation between frequency-integrated luminosity $L$, and the brightness temperature luminosity $L_{\mathrm{CO}}$ \citep[cfr. App. A, ][]{obreschkow2009}:
\begin{equation}
L_{\rm{CO}}=(8 \pi k_b)^{-1} \lambda_e^3 L = (8 \pi k_b)^{-1} \left(\frac{c}{\nu_e}\right)^3 L, 
\end{equation}
we obtain the L$_{\rm{CO}}$(6-5) luminosity in units of [$\rm{K\,\,km\,s^{-1}\,\,pc^2}$] and find that $L_{\rm{CO}}$(6-5)$= 1.58\,L_{\rm{CO}}$(1-0), while $L_{\rm{CO}}$(5-4)$= 1.75\,L_{\rm{CO}}$(1-0); all the results presented for the C0(6-5) transition also remain largely valid for the CO(5-4) transition.

This line luminosity can be converted into the line integrated flux, $S_{\rm{CO}}$ such that 
\begin{equation}\label{lco}
L_{\mathrm{CO}}=3.25\times 10^7 S_{\mathrm{CO}}\Delta V \nu_{obs}^{-2}(1+z)^{-3}D_{L}^2
\end{equation}
where $S_{\mathrm{CO}}\Delta V \equiv S^V$ is the velocity integrated flux [$\mathrm{Jy\,km\,s^{-1}}$], $\nu _{obs}$ is the observed frequency in GHz, and $D_L$ is the luminosity distance. We then assume that the CO(6-5) line has a gaussian profile with a width given by the rotational velocity of the galaxy, $v_r$, which we take to be equal to 1.5 times the halo rotation velocity \citep[see][]{dayal2009}. The average value of $v_r$ for LAEs at both the redshifts considered is of order of $200\,\rm{km/s}$.
The observable CO flux $S_{\rm{CO}}$ can then be expressed as $S_{\rm{CO}}\simeq S^V/v_r$.

\begin{figure}
\includegraphics[scale=0.45]{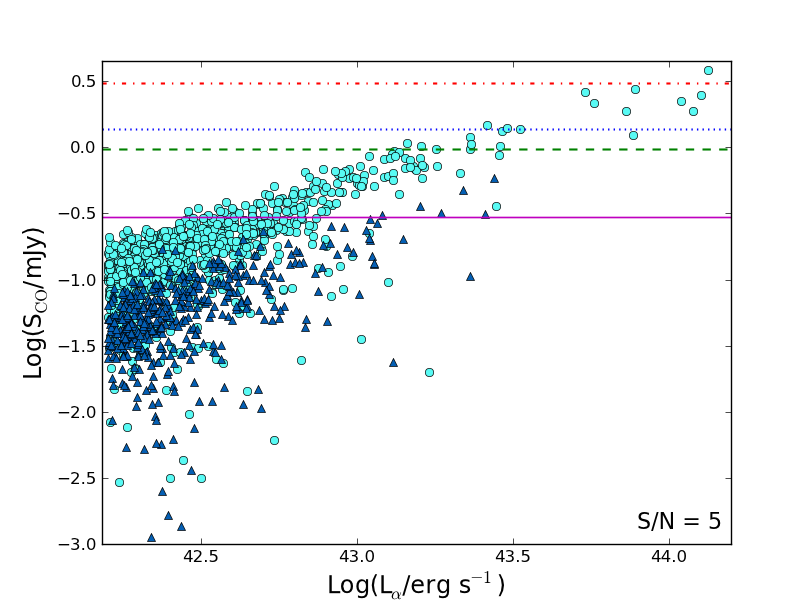}
\captionsetup{singlelinecheck=off}
\caption[]{The line integrated CO(6-5) flux, $S_{\rm{CO}}$, as a function of the $L_{\alpha}$ observed for LAEs at $z\approx5.7$ (circles) and $z\approx6.6$ (triangles). Horizontal lines represent the ALMA Cycle 0 sensitivity for a signal-to-noise ratio of 5, for an integration time of 1 hour (dot-dashed line), 5 hours (dotted line) and 10 hours (dashed line), assuming a spectral resolution of $50\,\,\rm{km/s}$. Solid line represents the sensitivity ($S/N=5$) achieved with 10 hours of integration time with 50 antennas.}
  \label{alma}
\end{figure}

As noted in Sec. \ref{co_detectability}, the CO(1-0) luminosity value scales with the observed Ly$\alpha$ luminosity; the line integrated flux is therefore also expected to behave in a similar way. This is indeed the case, as seen from Fig. \ref{alma}. Albeit with a large scatter, galaxies with the largest Ly$\alpha$ luminosity show the largest value of $S_{\rm{CO}}(6-5)$ at both $z \approx 5.7, 6.6$; as expected from a comparison of the CO(1-0) luminosities, the average value of $S_{\rm{CO}}$ is about 3 times lower at $z \approx 6.6$, compared to $z \approx 5.7$ (Tab. \ref{table2}). We find that at $z \approx 6.6$ none of the LAEs would be detectable in CO, even with an ALMA Cycle 0 integration time for about 10 hours. On the other hand, at $z \sim 5.7$, about 1-2\% of LAEs, i.e., those with $L_\alpha \geq 10^{43.2} {\rm erg \, s^{-1}}$ could be detectable with an ALMA integration time of 5-10 hours (i.e. a detection limit of $\sim$ 1-1.4 mJy respectively), assuming a signal-to-noise ratio $S/N=5$. However, using 50 antennas of ALMA with an integration time of 10 hours increases the sensitivity to about $0.3 \,\rm{mJy}$ making a significant change such that about 13\%, 1.4\% of LAEs become detectable at $z \approx 5.7$ and $6.6$ respectively.

\begin{figure}
\includegraphics[scale=0.45]{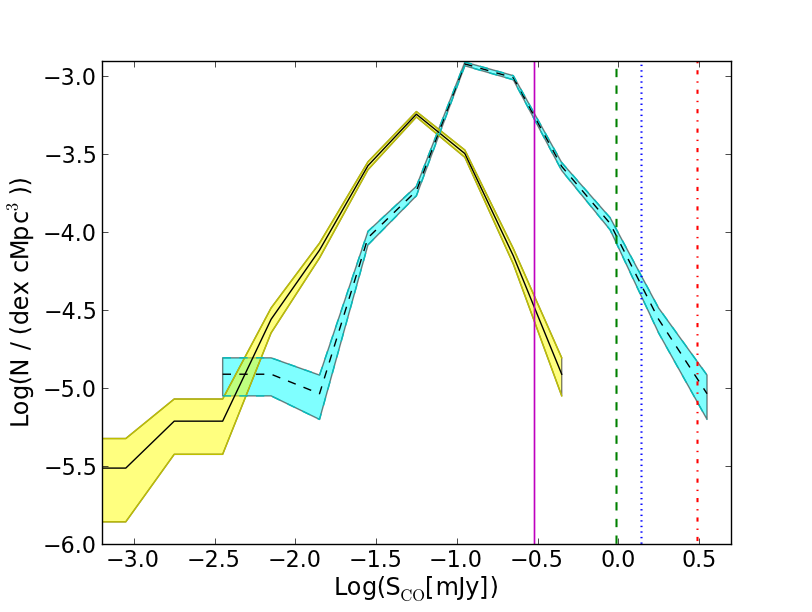}
\captionsetup{singlelinecheck=off}
\caption[]{The number density of LAEs at $z \approx 5.7$ and 6.6 are shown as a function of the line integrated CO(6-5) flux using dashed and solid lines respectively; shaded regions show the poissonian errors. Vertical dot-dashed, dotted and dashed lines represent the ALMA sensitivity limits for 16 antennas assuming a signal to noise ratio, $S/N=5$, for an integration time of 1, 5 and 10 hours respectively. Vertical solid line represents the sensitivity ($S/N=5$) achieved with 10 hours of integration time with 50 antennas.}
  \label{luminosityfunction}
\end{figure}

To clarify such results, we present the CO `flux function', i.e. the number density of LAEs as a function of the line integrated CO flux, $S_{\rm{CO}} (6-5)$. As is clearly seen again, none of the $z \approx 6.6$ LAEs are detectable with the Cycle 0 of ALMA, even for a 10 hour detection limit of about 1 mJy. On the other hand, the $z \approx 5.7$ flux function, extends into the 1-10 ALMA detection bands as mentioned above. However, the peak of the function lies at $S_{\rm{CO}} \approx 0.1$ mJy at $z \approx 5.7$; due to the lower $\rm{H_2}$ mass at $z \approx 6.6$, the flux function peaks at a lower value of about 0.06 mJy. As mentioned above, using 50 ALMA antennas for an integration time of 10 hours increases the number of LAEs detectable in CO; as expected, this leads to a larger part of the 'flux function' being observed at both the redshifts considered.

\section{Summary and Conclusions}
Using a physically motivated model, we present results concerning the ${\rm H_2}$ fraction, and the total ${\rm H_2}$ mass inside the ISM of LAEs at $z \approx 5.7$ and $6.6$ by coupling a semi-analytic model of the molecular content of galaxies, to a previously developed LAE model that reproduces a number of observed LAE data sets. 

We start by using cosmological SPH simulations to obtain the physical properties of each simulated galaxy, including the total halo/gas/stellar mass ($M_h/M_g/M_*$), the SFR ($\dot M_*$), the mass weighted gas/stellar metallicity ($Z_g/Z_*$), the mass weighted age ($t_*$) and the mass weighted gas temperature; the values of $Z_*, t_*, \dot M_*$ of each galaxy are then used to produce its intrinsic spectrum using {\tt STARBURST99} \citep{leitherer1999}. Assuming SNII to be the primary sources of dust, we calculate the ISM dust attenuation of both Ly$\alpha$ and continuum photons, and the IGM Ly$\alpha$ transmission, to select galaxies as LAEs \citep{dayal2009, dayal2010a, dayal2011b}. For each of these galaxies, the value of $f_{\rm H_2}$ is calculated using the analytic model proposed by \citet{krumholz2008, krumholz2009} and \citet{mckee2010}, which considers an idealized spherical molecular cloud immersed in a uniform and isotropic ${\rm H_2}$-dissociating Lyman-Werner radiation field; under approximations of steady state, the value of $f_{\rm H_2}$ is solely a function of the dust optical depth to LW photons, and a dimensionless parameter, $\chi$. The latter depends on: (a) the intensity of the LW field which is determined by $\dot M_*$, (b) the total dust mass, $M_d$, which both enables ${\rm H_2}$ formation on the surface of dust grains, and shields the $\rm{H_2}$ to LW photons, and (c) the cold \HI mass, $M_{\rm{CNM}}$, providing the key ingredient to produce ${\rm H_2}$.

Using this model, we find that at $z \approx 5.7$, the value of $f_{\rm H_2}$ peaks and ranges between $0.5-0.9$ for intermediate mass LAEs with $M_* \approx 10^{9-10} M_\odot$, decreasing for both smaller and larger galaxies; this trend also holds at $z \approx 6.6$. Such behaviour can be explained as follows: compared to intermediate mass galaxies, smaller galaxies have smaller SFR (and hence a lower value of the dissociating LW field), lower dust masses for forming ${\rm H_2}$, and a lower cold gas (and dust) mass for self-shielding the ${\rm H_2}$ so formed against the dissociating LW field; on the other hand, the larger SFR compared to the dust and cold gas mass in the largest galaxies leads to a lower $f_{\rm H_2}$. Such an argument is validated by the fact that at a given stellar mass, galaxies with the lowest SFR, largest cold gas mass, and largest dust mass have the largest value of $f_{\rm H_2}$. Further, we find that for LAEs with comparable SFR/gas mass/dust masses, $f_{\rm H_2}$ is about twice as large at $z \approx 5.7$ than at $6.6$; to reproduce the LAE UV luminosity function data, the dust distribution radius at $z \approx 6.6$ is about 1.6 times larger than that at 5.7, for similar LAE properties. As a result, the dust optical depth to LW photons is lower at $z \approx 6.6$ which leads to a larger amount of ${\rm H_2}$ being dissociated.

We then translate the $\rm{H_2}$ mass we obtain into a CO luminosity. Recently, \cite{wagg2009} used the green bank telescope to look for CO emission in two LAEs at $z > 6$ both of which resulted in non-detections. To compare our model predictions to their observations, we used a value of $\alpha=0.8\,\rm{M_{\odot}\, K \, km \, s^{-1} \, pc^2}$ to translate the ${\rm H_2}$ mass into a CO(1-0) luminosity, $L_{\rm{CO}}$. We find that the value of $L_{\rm{CO}}$ scales with $L_\alpha$, for both the redshifts considered; although the value of $f_{H2}$ is the largest for the intermediate mass/luminosity LAEs (see Sec. \ref{gal_prop}), the increasingly large neutral hydrogen mass, $M_{HI}$, in the star forming disk of increasingly massive galaxies wipes out such a subtle signature. This result also implies that the brightest LAEs are the best candidates for molecular emission searches. As a reasonable validation of our model, our theoretical $L_{\rm{CO}}$ estimates are in accord with the upper limits found by \cite{wagg2009} in their experimental work.

At $z \approx 5.7, 6.6$, the lowest CO rotational transition observable with ALMA is the CO(6-5). We find that at $z \approx 5.7$, about 1-2\% of the LAEs, i.e. those with $L_\alpha \geq 10^{43.2} {\rm erg \, s^{-1}}$, could be detectable with an integration time of 5-10 hours respectively, assuming a signal-to-noise ratio $S/N=5$. Our results at $z \approx 6.6$ are more pessimistic; none of the LAEs would be detectable in CO, even with an ALMA integration time of about 10 hours. We also present the CO `flux function', the number density of LAEs as a function of $S_{\rm{CO}}$ where we show that the number density of objects peaks at a value of about 0.1 mJy, which is much beyond the sensitivity of ALMA; this peak shifts to progressively lower values with increasing redshift.

Finally, we discuss the main caveats in the model. First, the calculations presented here concern only average quantities in a spherically symmetric framework and a full calculation of the radial dependence of $f_{\rm H_2}$ is the subject of an ongoing work. Secondly, the dust masses used in this work have been calculated assuming SNII to be the primary sources of dust production which is a reasonable assumption given that a number of authors \citep{todini2001, dwek2007} have shown that the contribution of AGB stars becomes progressively less important and at some point negligible towards higher redshifts ($z \gsim 5.7$) when the Universe is less than 1 Gyr old. However, it must be noted that under certain conditions thought to hold in quasars, in which extremely massive starbursts occur, the contribution of AGB can become important somewhat earlier \citep[see][]{valiante2009}. Thirdly, the stellar distribution scale (and radial extent of the MC) is based on estimates following the results of \citet{bolton2008}, who have derived fitting formulae relating the V-band luminosity and the stellar distribution scale from their observations of massive, early type galaxies between $z=0.06-0.36$; however, we note that such estimates are in surprisingly good agreement (to within 1$-\sigma$) with recent observational results of $z \approx 5.7$ LAEs \citep{malhotra2011}. Fourthly, the large cosmological volume simulated naturally results in a low mass resolution, such that we are unable to resolve density and temperature of the gas in the interior of individual galaxies. We have therefore estimated the ISM gas temperature distribution over broad halo mass bins. Further, in our calculations, we have assumed that half of the gas mass with temperatures of $T<10^4\,\rm{K}$ is cold, with $T<100\,\rm{K}$. Finally, we have used a ULIRG value for the factor $\alpha$, used to translate the $\rm{H_2}$ mass into a CO luminosity. However, this parameter depends on a number of poorly known properties of high-$z$ galaxies, such as the spatial/mass distribution of their MCs and the gas metallicity \citep{leroy2009, narayanan2011, genzel2011}. It is hoped that upcoming data from state of the art instruments such as ALMA will be able to shed light and clarify such thorny issues.

\section*{Acknowledgements}
We thank S. Borgani, L. Tornatore and A. Saro for providing the simulations used in this work. We acknowledge the DAVID Workshop VI for numerous discussions about this work.
LV thanks Raffaella Schneider for insightful suggestions and comments, PD warmly thanks for hospitality during the early phases of this research.

\bibliographystyle{mn2e}
\bibliography{mh}

\end{document}